\documentclass[conference]{IEEEtran}
\IEEEoverridecommandlockouts
\usepackage{cite}
\usepackage[main=english]{babel}
\usepackage{amsmath,amssymb,amsfonts,amstext,amsthm}
\usepackage{algorithmic}
\usepackage{graphicx}
\usepackage{textcomp}
\usepackage{xcolor}

\usepackage[flushleft]{threeparttable}
\usepackage{hyperref}
\usepackage{multirow}
\usepackage[bottom]{footmisc}
\usepackage[utf8]{inputenc}

\def\BibTeX{{\rm B\kern-.05em{\sc i\kern-.025em b}\kern-.08em
    T\kern-.1667em\lower.7ex\hbox{E}\kern-.125emX}}
\begin{document}

\title{SSSD-ECG-nle: New Label Embeddings with Structured State-Space Models for ECG generation\\
{\footnotesize \textsuperscript{}}
\thanks{*corresponding author}
}

\author{
\IEEEauthorblockN{1\textsuperscript{st} Sergey Skorik \textsuperscript{*}}
\IEEEauthorblockA{\textit{ISP RAS}\\
Moscow, Russia \\
skorik@ispras.ru}
\and
\IEEEauthorblockN{2\textsuperscript{nd} Aram Avetisyan}
\IEEEauthorblockA{\textit{ISP RAS}\\
Moscow, Russia \\
a.a.avetisyan@ispras.ru}
}

\maketitle

\begin{abstract}
An electrocardiogram (ECG) is vital for identifying cardiac diseases, offering crucial insights for diagnosing heart conditions and informing potentially life-saving treatments. However, like other types of medical data, ECGs are subject to privacy concerns when distributed and analyzed. Diffusion models have made significant progress in recent years, creating the possibility for synthesizing data comparable to the real one and allowing their widespread adoption without privacy concerns. In this paper, we use diffusion models with structured state spaces for generating digital 10-second 12-lead ECG signals. We propose the SSSD-ECG-nle architecture based on SSSD-ECG with a modified conditioning mechanism and demonstrate its efficiency on downstream tasks. We conduct quantitative and qualitative evaluations, including analyzing convergence speed, the impact of adding positive samples, and assessment with physicians' expert knowledge. Finally, we share the results of physician evaluations and also make synthetic data available to ensure the reproducibility of the experiments described.

\end{abstract}


\section{Introduction}

An Electrocardiogram (ECG) is a non-invasive diagnostic tool that assesses the overall cardiac health of a patient by recording and analysing the temporal changes in the heart's electrical potential. It serves as the primary method for diagnosing cardiovascular diseases. ECG analysis is a labour-intensive process and the growing availability of digital ECG signals leads to overwork of medical staff, which increases the chance of medical errors \cite{b1}. One way to address this involves the application of automated interpretation methods. Deep neural networks (DNNs) are among the most popular and effective methods for analysing ECG data \cite{b2}. Recent studies have demonstrated outstanding results in the task of cardiovascular disease detection using DNNs \cite{b3,b4}. One of the key aspects of effectively utilising DNNs is the requirement for a large dataset of high-quality and well-documented data. Most studies use open ECG datasets \cite{b5,b6,b7,b8}, which are significantly smaller than private analogues, which leads to a decrease in the robustness and generalizability of the model \cite{b9,b10,b11}. 

One of the most significant challenges standing in the way of the widespread adoption of ECG in open sources is data privacy, which is particularly sensitive in the healthcare domain. Common methods for ensuring privacy include federated learning, differential privacy, etc. \cite{b12,b13}. However, these approaches are vulnerable to various types of attacks, complicating their implementation. Another direction in solving privacy issues is data augmentation, although it is limited in application methods.

In recent years, generative models have significantly advanced in data synthesis tasks, achieving quality comparable to real data \cite{b14}. This progress enables their use as a method for protecting personal data, where the model takes real samples as input, trains on them, and produces synthetic anonymized examples. However, for reliable protection, the generative model must be objective and not biased towards the training set \cite{b15}. This requires extensive model evaluation from both a quantitative and qualitative perspective. 

In this study, we focus on generating 10-second digital 12-lead ECGs, which are most commonly used. We employ a diffusion neural network based on the SSSD architecture \cite{b16}, which has proven effectiveness in ECG synthesis tasks \cite{b17}. Our contribution includes:

\begin{figure*}[htp]
\centerline{\includegraphics[width=0.9\textwidth]{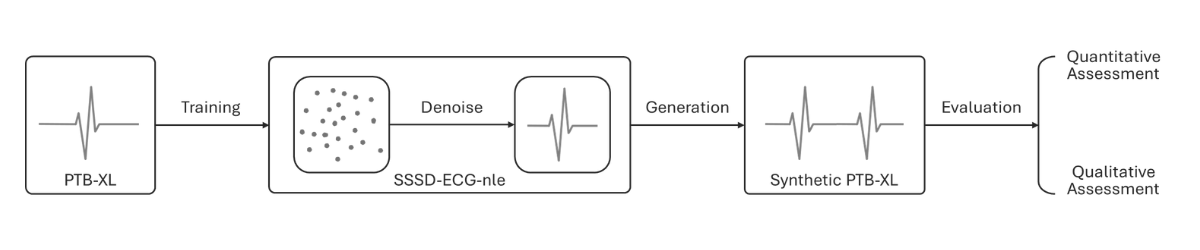}}
\caption{The overall pipeline of ECG generation. First, we train SSSD-ECG-nle model on train dataset of 12-lead ECG records. Then, we generate synthetic copy of the training data. Finally, we evaluate the generated samples both qualitatively and quantitatively.}
\label{fig:Methodology}
\end{figure*}

\begin{itemize}
    \item We modified the conditioning mechanism in SSSD-ECG, encouraging the model to generate neutral examples and enhancing the consistency of real and synthetic data in terms of metrics on downstream task. With new label embeddings (nle) we named the proposed architecture variant SSSD-ECG-nle. 
    \item We conducted extensive quantitative and qualitative evaluation of the generative model. The quantitative analysis includes assessing synthetic data using Train Synthetic - Test Real (TSTR) and Train Real - Test Synthetic (TRTS) metrics. Also it includes analyzing convergence speed and the impact of adding positive samples. The qualitative analysis involves a questionnaire for physicians, asking them to write a text report for an ECG and answer the question: "Is this a generated record?". This examination, based on approximately 400 samples, gives comprehensive evaluation of the quality of the generated synthetic samples.
    \item We share\footnote{\url{https://github.com/ispras/EcgLib}} the code with synthetic data for the reproducibility of the described experiments and facilitate further research.
\end{itemize}

\begin{figure*}[t]
\centering
\includegraphics[width=0.9\textwidth]{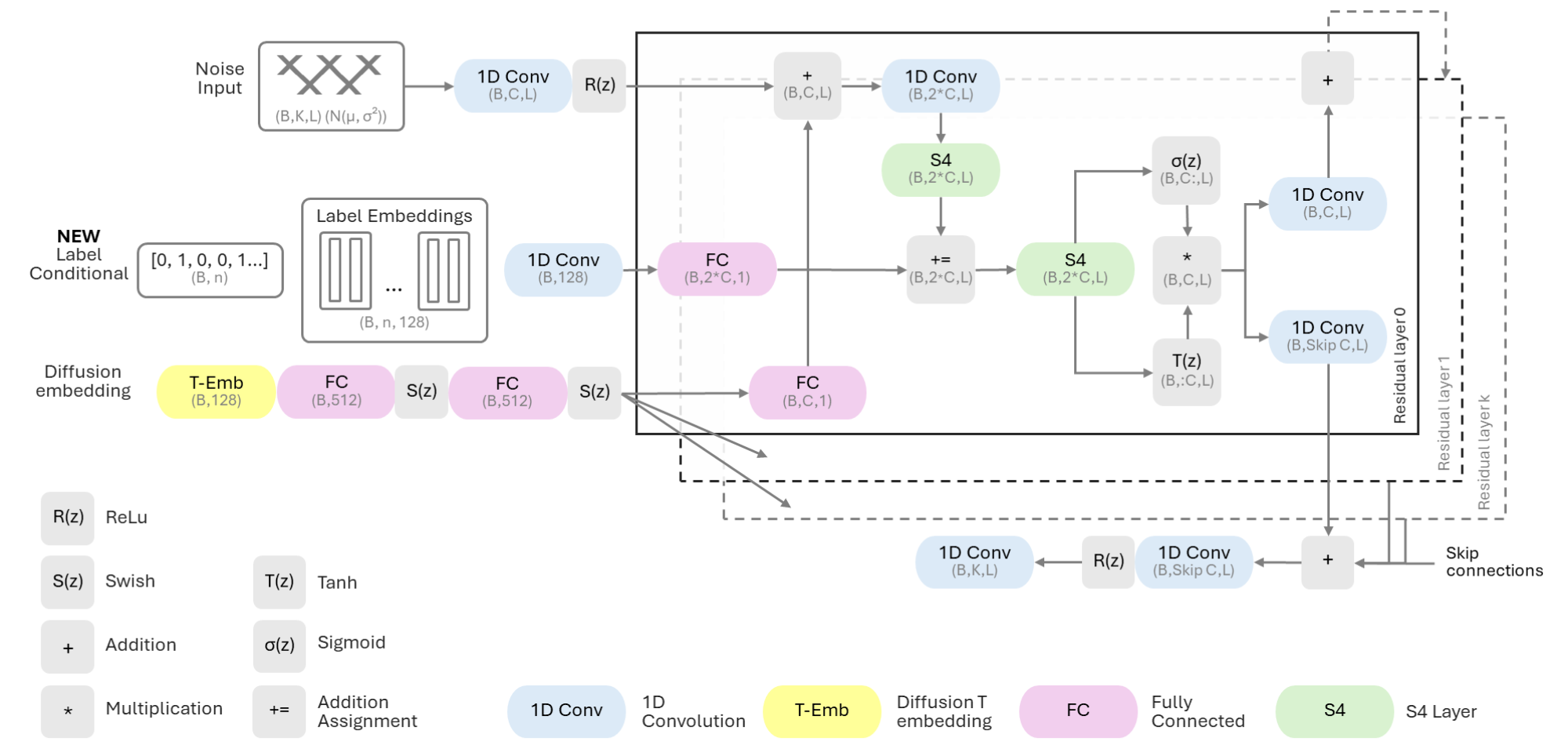}
\caption{Schematic representation of the SSSD-ECG-nle model architecture.}
\label{fig:SSSD_ECG_nle}
\end{figure*}

\section{Related Work}

Generative models for synthesizing ECG records are generally categorized into three types: Variational Autoencoders (VAE), Generative Adversarial Networks (GAN), and Denoising Diffusion Probabilistic Models (DDPM) \cite{b18}.

A significant amount of work in the task of ECG generation primarily employs the GAN models \cite{b19,b20,b21,b22}. However, the adversarial nature of the training process often leads to instability \cite{b23}. This instability is further complicated by the specifics of ECG data: the limited amount of samples encourages preprocessing steps and segmentation of the signal into individual beats, which complicates reproducibility and the consistency of the results.

Despite the popularity of GANs, the application of DDPM for ECG synthesis has significantly increased over the past year \cite{b24,b25,b26}. This reflects the overall trend towards using diffusion models across various domains \cite{b27}. The more stable training process of DDPMs allows feeding the entire signal without preprocessing, thereby enhancing the quality and diversity of the generated samples \cite{b17}, \cite{b25}. In a conditional setup, this approach enables the generation of a complete dataset and quantitative evaluation of the diffusion model using informative metrics such as train synthetic - test real (TSTR) and train real - test synthetic (TRTS) \cite{b28}.

Our research is inspired by SSSD-ECG \cite{b17}, a significant advancement in the conditional generation of digital 10-second ECGs. With the widespread availablity of digital 12-lead ECG data, we recognize that focusing on this primary domain improves the overall quality of generated samples and excludes various preprocessing steps, such as converting the signal into a spectrogram and back \cite{b26}.

Our goal is to refine the conditioning mechanism of SSSD-ECG by encouraging neutral examples. The validation and evaluation of generative models are subjects of discussion. These areas are particularly sensitive in fields that require high-quality expert assessment. Therefore, we quantitatively and qualitatively evaluate the generative model, considering issues of convergence and the interpretability of the obtained synthetic samples.

\section{Background}

\subsection{Denoising diffusion probabilistic models}
Several studies \cite{b29}, \cite{b30}, \cite{b31} proposed similar ideas that form the foundation of the DDPM. These ideas can be summarized into forward and backward diffusion processes. The forward diffusion process involves sequentially applying white noise $\boldsymbol\varepsilon \sim \mathcal{N}(\mathbf{0}, \mathbf{I})$ to the original sample $\boldsymbol x \sim \pi(\boldsymbol x)$ over $T$ steps with a transition kernel $q(\boldsymbol x_t | \boldsymbol x_{t-1}) = \mathcal{N}(\boldsymbol x_t | \sqrt{1 - \beta_t}\boldsymbol x_{t-1}, \beta_t\mathbf{I})$ where $\beta_t \in (0, 1)$. It can be represented in a closed-form without trained parameters $q(\boldsymbol x_t | \boldsymbol x_{0}) = \mathcal{N}(\boldsymbol x_t | \sqrt{\bar{\alpha}_t} \boldsymbol x_{0}, (1 - \bar{\alpha}_t)\mathbf{I})$, $\bar{\alpha}_t = \prod_{s=1}^t(1 - \beta_s)$. The reverse process begins from point $\boldsymbol x_T \sim \mathcal{N}(\boldsymbol 0, \mathbf{I})$ and sequentially denoises the sample  using a neural network that learns reverse transition $p(\boldsymbol x_{t-1} | \boldsymbol x_t, \boldsymbol \theta) = \mathcal{N}(\boldsymbol x_{t-1}| \boldsymbol \mu(\boldsymbol x_t, \boldsymbol \theta, t), \Tilde{\beta}_t\mathbf{I})$, $\Tilde{\beta}_t = \frac{(1 - \bar{\alpha}_{t-1})}{1 - \bar{\alpha}_{t}}\beta_t$. It was demonstrated \cite{b31} that by using a specific type of $\boldsymbol \mu(\boldsymbol x_t, \boldsymbol \theta, t)$, that attempts to predict random noise $\boldsymbol\varepsilon_t$, objective loss $\mathcal{L} = \sum_{t=1}^T \mathcal{L}_t$ can be expressed as

\begin{equation}
    \mathcal{L}_t = \mathbb{E}_{t, \boldsymbol x_0, \boldsymbol\varepsilon} \left[\|\boldsymbol\varepsilon -  \boldsymbol\varepsilon(\sqrt{\bar{\alpha}_t} \boldsymbol x_{0} + \sqrt{1 - \bar{\alpha}_t}\boldsymbol\varepsilon, \boldsymbol\theta, t) \|^2_2\right],
\end{equation}
where $t \sim \mathcal{U}(\{1, T\})$, $\boldsymbol x_0 \sim \pi(\boldsymbol x)$ and $\boldsymbol\varepsilon \sim \mathcal{N}(\mathbf{0}, \mathbf{I})$.

\subsection{Structured state space models}

The theory of Structured State Spaces (S4) evolves through three consecutive works \cite{b32,b32,b34}. These works aim to reconstruct time series with long-range dependencies. The first study \cite{b32} formalizes the concept of optimal reconstruction by introducing the $hippo[u](T)$ operator, which, for a given $u(t)\big|_{t \leqslant T}$, obtains a vector of coefficients $\boldsymbol x(t) \in \mathbb{R}^N$ of an orthogonal polynomial $g^{(t)}$ that best approximates the original signal $u(t)$ with respect to the measure $\mu$ supported on $(-\infty, T]$. In \cite{b32}, it is shown that state dynamics, in vector form, can be viewed as:

\begin{equation}
    \Dot{\boldsymbol x}(t) = \boldsymbol A \boldsymbol x(t) + \boldsymbol B u(t).
\end{equation}

Thus, the reconstruction of signal u(t) requires $N$ coefficients of the polynomial $g^{(t)}$, which is computationally infeasible in the case of DNNs with large input dimensionality. The following work \cite{b33} introduces an additional equation that involves a linear combination of states and skip connection. Consequently, the original signal $u(t)$ is transformed to $y(t)$ in a sequence-to-sequence manner through the state space representation

\begin{equation}\label{S4}
    \begin{cases}
    \Dot{\boldsymbol x}(t) = \boldsymbol A \boldsymbol x(t) + \boldsymbol B u(t) \\
    y(t) = \boldsymbol C \boldsymbol x(t) + \boldsymbol D u(t).
    \end{cases}
\end{equation}

In \cite{b33}, it was shown that by restrcting $\boldsymbol A$ to a certain class of matrices, we always obtain a $hippo$-interpretation of the state space. Therefore, by training $\boldsymbol A$ within this class, we will select the best reconstruction of $u(t)$ according to theory described in \cite{b32}. The research detailed in \cite{b34} addresses computational challenges by structuring $\boldsymbol A$ as Normal Plus Low-Rank (NPLR) matrices, obtaining structured state spaces \eqref{S4} with efficient computation and reconstruction.

\section{Methods}

We closely follow the methodology in \cite{b17} to ensure our results are consistent and reproducible. Overall pipeline can be viewed at Figure \ref{fig:Methodology}.

\subsection{Dataset}

We utilize the publicly available PTB-XL dataset \cite{b5}, which consists of 21837 ECG records. Each ECG is a digital 12-lead, 10 seconds ECG record. The ECGs have a sampling frequency of either 500 Hz or 100 Hz. For this study, we opted to work with a sampling frequency of 100Hz to reduce the computational complexity of generation. Additionally, we use only 8 linearly independent leads for model training, and subsequently reconstruct the remaining 4 leads in a manner similar to \cite{b17}. All the ECGs have been preprocessed with a bandpass filter. We do not employ additional preprocessing and maintain the original amplitude range of the ECGs, as we observed that the trained model is sensitive to such changes.

\subsection{SSSD-ECG limitations}\label{methods:sssd_ecg_limitations}
The SSSD-ECG architecture proposed in \cite{b17} is based on SSSD diffusion model \cite{b16} and adapt conditioning mechanism to a multilabel generation. It replaces the binary imputation mask with the corresponding vector of labels.  However, the proposed conditioning has several limitations.

The primary limitation lies in the mechanism for obtaining embeddings from the \textit{label vector}. In a multilabel setup, the vector $\boldsymbol y = [0, 1, 0, 0, 1, \ldots]$ has length $N$ where each position $y_i \in \{0, 1\}$ indicates the presence or absence of disease number $i$. To convert $\boldsymbol y$ to embeddings $\boldsymbol e$ the authors use matrix multiplication $\boldsymbol y \times \boldsymbol E$ where $\boldsymbol E =  \|\boldsymbol e_i\|_{i=1}^N$ is a \textit{Label Embeddings} matrix, $\boldsymbol e_i \in \mathbb{R}^{128}$. To illustrate the limitation of this multiplication consider the binary classification scenario, i.e., $N=1$, $y \in \{0, 1\}$ and $\boldsymbol E = \boldsymbol e_1$. Thus

\begin{equation}\label{eq:sssd_ecg_multiplication}
    \underbrace{\boldsymbol y}_{1\times 1} \times \underbrace{\boldsymbol E}_{1\times 128} = y \ast \boldsymbol e_1,
\end{equation}
where $\ast$ denotes element-wise multiplication. Intuitively, it seems that in the binary scenario, we should consider 2 embeddings, describing $y_i=0$ and $y_i=1$. However, matrix multiplication gives us $\boldsymbol e_1$ for $y_i=1$ and $0 \ast \boldsymbol e_1 = \boldsymbol 0$ for $y_j=0$, failing to convey information about neutral examples without observed abnormalities. In Table \ref{conditioning_comparsion}, we demonstrate how this affects the quality of generated neutral samples.

Another limitation is the absence of a padding index in the matrix multiplication process, which could be useful in setups like classifier-free guidance \cite{b35}.

\subsection{Generative model}\label{methods:gen_model}

\begin{figure*}[ht]
\centering
\includegraphics[width=0.9\textwidth]{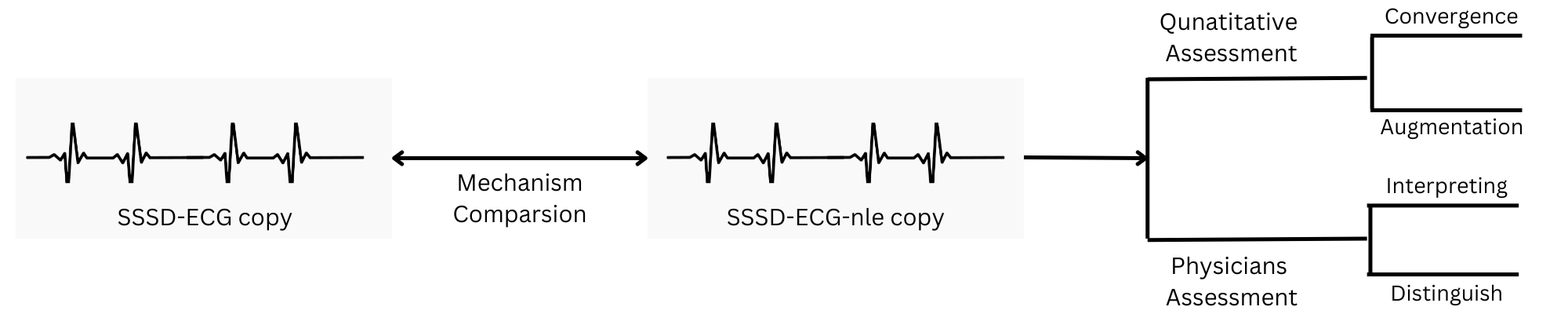}
\caption{Schematic representation of the assessment pipeline.}
\label{fig:assessment_pipeline}
\end{figure*}

To address these issues, we reconsider the conditioning mechanism. We modify the matrix $\boldsymbol E$ by resizing it to $(N, 2, 128)$ so that for each label $y_{i}$, the corresponding embedding $e_{iy_{i}}$ can be selected similarly to the multiclass scenario as if from a table. This approach allows us to obtain $N$ embeddings of dimension $128$. To bring them back to the original size without changing the residual layers, we add a trainable convolution with a kernel size of $1 \times 1$ to fold the $N$ channels into $1$. We have named the proposed architecture variant SSSD-ECG-nle. The structure of this architecture is presented in Figure \ref{fig:SSSD_ECG_nle}.

\subsection{Generative model evaluation}\label{methods:model_eval}

Evaluating generative models is less straightforward than evaluating discriminative models, making it a topic of separate discussion. There are studies aimed at systematizing evaluation in the time-series domain \cite{b36},\cite{b28}. Evaluation metrics for generative models can be broadly categorized into three types: relation, visualization, and downstream.

Relation metrics quantify the relationship between synthetic and real samples. This includes metrics like Euclidean distance and dynamic time warping to measure the relationship between specific samples, as well as Wasserstein distance, maximum mean discrepancy, and others for assessing the relationship between sample distributions. Visualization metrics provide a way to visually inspect patterns and differences between synthetic and real samples. Given that ECGs are medical data, visual evaluation of generative models is challenging and typically requires expert knowledge. Apart from direct visual evaluation, feature representation techniques such as t-SNE \cite{b37} can be used. To quantify visual evaluations, indirect tasks can be set for physicians, such as determining whether a sample is synthetic, followed by evaluating the accuracy of their classifications. Downstream metrics assess the effectiveness of using synthetic data for the downstream task. Common metrics include TSTR, TRTS, and TSTS.

In our work, we do not use relation metrics, as they proved to be uninformative during experiments. For visualization metrics, we complied two datasets and asked physicians with assessing each. A detailed description is provided in Section \ref{exp:qualitative_assessment} along with a quality assessment. In downstream metrics, we used all metrics except TSTS, as it also proved to be uninformative during preliminary experiments. Typically, models under this metric perfectly solve the task, resulting in scores close to 1.

\subsection{Downstream task}

For the downstream task, we focus on predicting ECG statements. We split the dataset into 10 folds, where the first eight are used for training, the ninth serves as validation, and the tenth as test. We opt for binary classification rather than multilabel classification as the primary method because this approach allows us to better distinguish specific abnormalities using an ensemble of models \cite{b10}. Furthermore, the test dataset includes multiple diseases that appear in fewer than 10 samples, complicating objective evaluation. We employ Xresnet1d50, a 1D adaptation of Xresnet \cite{b38}, for training. The quality of a medical diagnostic algorithm is primarily indicated by its \textit{Sensitivity}, which measures the ability to identify sick individuals. Following an automatic diagnosis, an expert validates the results. Inaccurate algorithm responses increase the burden on doctors due to additional checks. So the \textit{Precision} and \textit{Specificity} also gives a significant information. To evaluate our models, we select G-mean $= \sqrt{Sens. \cdot Spec.}$ and F1-score instead of macro-averaged ROC-AUC, as these metrics gives an aggregated view of the classifier’s performance in terms of the three described metrics. Other hyperparameters, such as learning rate, batch size, and weight decay, are configured and tuned similarly to those in \cite{b17}.

\section{Experiments}

For evaluating our generative model, we implement the following experimental pipeline: Initially, we train the SSSD-ECG-nle model on the first 8 folds of the PTB-XL dataset, providing it with contextual information about the presence or absence of conditioned diseases within each signal. Subsequently, we generate a synthetic copy of the PTB-XL dataset. To achieve this, we extract the context from each real signal to guide the diffusion model, then generate a corresponding synthetic signal, and match it with the corresponding real record. A schematic representation of assessment pipeline is shown in the Figure \ref{fig:assessment_pipeline}.

\begin{table*}[htbp]
\caption{Conditioning mechanism comparsion}
\begin{threeparttable}
\centering
\resizebox{\textwidth}{!}{
\begin{tabular}{|c|cccccccc|}
\hline
\textbf{Dataset }                & \multicolumn{4}{c|}{\textbf{Train Synthetic dataset -- Test Real dataset (TSTR)}}                                                                                                                                                                                                                                                                                                                     & \multicolumn{4}{c|}{\textbf{Train Real dataset -- Test Synthetic dataset (TRTS)}}                                                                                                                                                                                                                                                                                                         \\ \hline
\multirow{2}{*}{\textbf{}}        & \multicolumn{8}{c|}{\textbf{Abnormality}}                                                                                                                                                                                                                                                                                                                                                                                                                                                                                                                                                                                                                                                                                                                                                  \\ \cline{2-9} 
                                  & \multicolumn{1}{c|}{\textbf{\begin{tabular}[c]{@{}c@{}}AFIB\\ G-mean\quad f1-score\end{tabular}}} & \multicolumn{1}{c|}{\textbf{\begin{tabular}[c]{@{}c@{}}1AVB\\ G-mean\quad f1-score\end{tabular}}} & \multicolumn{1}{c|}{\textbf{\begin{tabular}[c]{@{}c@{}}PVC\\ G-mean\quad f1-score\end{tabular}}} & \multicolumn{1}{c|}{\textbf{\begin{tabular}[c]{@{}c@{}}CRBBB\\ G-mean\quad f1-score\end{tabular}}} & \multicolumn{1}{c|}{\textbf{\begin{tabular}[c]{@{}c@{}}AFIB\\ G-mean\quad f1-score\end{tabular}}} & \multicolumn{1}{c|}{\textbf{\begin{tabular}[c]{@{}c@{}}1AVB\\ G-mean\quad f1-score\end{tabular}}} & \multicolumn{1}{c|}{\textbf{\begin{tabular}[c]{@{}c@{}}PVC\\ G-mean\quad f1-score\end{tabular}}} & \textbf{\begin{tabular}[c]{@{}c@{}}CRBBB\\ G-mean\quad f1-score\end{tabular}} \\ \hline
\multicolumn{1}{|c|}{SSSD-ECG}    & \multicolumn{1}{c|}{0.848\quad \textbf{0.720}}                                                              & \multicolumn{1}{c|}{0.860\quad 0.270}                                                             & \multicolumn{1}{c|}{\textbf{0.872}\quad 0.338}                                                            & \multicolumn{1}{c|}{0.931\quad 0.474}                                                              & \multicolumn{1}{c|}{0.904\quad 0.450}                                                              & \multicolumn{1}{c|}{0.867\quad 0.109}                                                             & \multicolumn{1}{c|}{0.834\quad 0.187}                                                            & 0.943\quad 0.318                                                              \\ \hline
\multicolumn{1}{|c|}{SSSD-ECG-nle} & \multicolumn{1}{c|}{\textbf{0.922}\tnote{b}\quad 0.709}                                                    & \multicolumn{1}{c|}{\textbf{0.871}\quad \textbf{0.293}}                                                    & \multicolumn{1}{c|}{0.850\quad \textbf{0.561}}                                                   & \multicolumn{1}{c|}{\textbf{0.940}\quad \textbf{0.640}}                                                     & \multicolumn{1}{c|}{\textbf{0.985}\quad \textbf{0.858}}                                                    & \multicolumn{1}{c|}{\textbf{0.933}\quad \textbf{0.824}}                                                    & \multicolumn{1}{c|}{\textbf{0.864}\quad \textbf{0.649}}                                                   & \textbf{1.000}\quad \textbf{0.981}                                                     \\ \hline
\multicolumn{1}{|c|}{Real \tnote{a}}        & \multicolumn{1}{c|}{0.950\quad 0.744}                                                             & \multicolumn{1}{c|}{0.910\quad 0.391}                                                             & \multicolumn{1}{c|}{0.971\quad 0.820}                                                            & \multicolumn{1}{c|}{0.977\quad 0.642}                                                              & \multicolumn{1}{c|}{0.950\quad 0.744}                                                             & \multicolumn{1}{c|}{0.910\quad 0.391}                                                             & \multicolumn{1}{c|}{0.971\quad 0.820}                                                            & 0.977\quad 0.642                                                              \\ \hline
\end{tabular}
}
\label{conditioning_comparsion}
\begin{tablenotes}
    \item[a] We included a real dataset in the comparison to illustrate the gap between TSTR and TRTR metrics.
    \item[b] The best scores between synthetic datasets are represented by bold values.
\end{tablenotes}
\end{threeparttable}
\end{table*}

\subsection{Quantitative assessment}

\subsubsection{Condition mechanism refining}

In section \ref{methods:gen_model}, we have highlighted potential issues arising from the conditioning mechanism. Table \ref{conditioning_comparsion} presents a comparison of two approaches. The evaluation of the quality of generated samples is conducted using the TSTR and TRTS metrics. The primary observation reveals a substantial increase in the f1-score within the TRTS scenario. This underscores the principal limitation of the conditioning mechanism in SSSD-ECG: the embedding matrix from conditioning does not affect neutral samples. Consequently, a model trained on real data yields a high number of false positives when evaluated on a synthetic dataset. Additionally, we also observe an improvement in the TSTR metrics for all considered pathologies.

Another observation is the decrease in the TSTR metric for Premature Ventricular Complexes (PVC) compared to Complete Right Bundle Branch Block (CRBBB) and Atrial Fibrillation (AFIB), where the TRTR metric similarly and accurately distinguishes samples. Similar decreases were observed for other diseases dsuch as Inferior Myocardial Infarction (IMI), Complete Left Bundle Branch Block (CLBBB), Premature Atrial Complex (PAC), and others during preliminary experiments. We attribute this to the internal complexity of the considered class, convergence of the generative model \ref{exp:convergence}, and the volume of positive samples in the PTB-XL dataset. These and other challenges are addressed in section \ref{Conclusion}, indicating potential for future work.

Direct comparison under the same conditions seems essential to us. However, we encountered challenges with the reproducibility of the obtained TSTR metric \cite{b17} in a multilabel setup. Therefore, we report TSTR and TRTS results without redard to the work \cite{b17}. To assess the quality on downstream tasks, we employ weighted G-mean and f1-score, which consider the class imbalance in the dataset. Additionally, we report ROC-AUC to provide a consistent evaluation with \cite{b17}, \cite{b25}. The results are presented in Table \ref{multilabel_results}. 

\begin{table}[htbp]
\caption{Downstream metrics in multilabel classification}
\centering
\resizebox{0.48\textwidth}{!}{
\begin{tabular}{|ccc|ccc|}
\hline
\multicolumn{3}{|c|}{\textbf{TSTR}}                                                               & \multicolumn{3}{c|}{\textbf{TRTS}}                                                               \\ \hline
\multicolumn{1}{|c|}{\textbf{G-mean}} & \multicolumn{1}{c|}{\textbf{f1-score}} & \textbf{ROC-AUC} & \multicolumn{1}{c|}{\textbf{G-mean}} & \multicolumn{1}{c|}{\textbf{f1-score}} & \textbf{ROC-AUC} \\ \hline
\multicolumn{1}{|c|}{0.674}           & \multicolumn{1}{c|}{0.456}             & 0.820            & \multicolumn{1}{c|}{0.775}             & \multicolumn{1}{c|}{0.567}               & \multicolumn{1}{c|}{0.894}               \\ \hline
\end{tabular}
}
\label{multilabel_results}
\end{table}

\subsubsection{Real dataset augmenting}

Another approach to assess performance on downstream tasks is to consider the generated signals as augmentation of the original dataset and add a positive class to the training set. We compare this augmentation strategy with training on the original dataset (baseline), as well as training on a dataset where the number of positive samples is doubled through replication (double). The latter is necessary to eliminate the possibility of improvement solely through dataset enlargement. The comparison results are presented in Table \ref{dataset_augmentation}.

\begin{table*}[htbp]
\caption{Augmentation}
\begin{threeparttable}
\centering
\resizebox{\textwidth}{!}{
\begin{tabular}{|c|cccccccccclccll|}
\hline
\textbf{Metrics}           & \multicolumn{15}{c|}{\textbf{Abnormality}}                                                                                                                                                                                                                                                                                                                                                                                                                                                                     \\ \hline
\multirow{2}{*}{\textbf{}} & \multicolumn{3}{c|}{\textbf{AFIB}}                                                             & \multicolumn{3}{c|}{\textbf{CRBBB}}                                                               & \multicolumn{3}{c|}{\textbf{1AVB}}                                                                    & \multicolumn{3}{c|}{\textbf{CLBBB}}                                                            & \multicolumn{3}{c|}{\textbf{PVC}}                                                              \\ \cline{2-16} 
                           & \multicolumn{1}{c|}{baseline} & \multicolumn{1}{c|}{double} & \multicolumn{1}{c|}{synth. aug.\tnote{a}} & \multicolumn{1}{c|}{baseline} & \multicolumn{1}{c|}{double} & \multicolumn{1}{c|}{synth. aug.}    & \multicolumn{1}{c|}{baseline} & \multicolumn{1}{c|}{double}     & \multicolumn{1}{c|}{synth. aug.}    & \multicolumn{1}{c|}{baseline} & \multicolumn{1}{c|}{double} & \multicolumn{1}{c|}{synth. aug.} & \multicolumn{1}{c|}{baseline} & \multicolumn{1}{c|}{double} & \multicolumn{1}{c|}{synth. aug.} \\ \hline
G-mean                     & \multicolumn{1}{c|}{0.950}    & \multicolumn{1}{c|}{\textbf{0.954}}      & \multicolumn{1}{c|}{0.951}       & \multicolumn{1}{c|}{0.977}    & \multicolumn{1}{c|}{0.984}  & \multicolumn{1}{c|}{\textbf{0.985}} & \multicolumn{1}{c|}{0.910}    & \multicolumn{1}{c|}{\textbf{0.921}} & \multicolumn{1}{c|}{0.912}          & \multicolumn{1}{c|}{\textbf{0.976}}    & \multicolumn{1}{c|}{0.974}       & \multicolumn{1}{c|}{0.970}       & \multicolumn{1}{c|}{\textbf{0.971}}    & \multicolumn{1}{c|}{0.965}       &  \multicolumn{1}{c|}{0.948}                                \\ \hline
f1-score                   & \multicolumn{1}{c|}{0.734}    & \multicolumn{1}{c|}{0.796}      & \multicolumn{1}{c|}{\textbf{0.806}}       & \multicolumn{1}{c|}{0.642}    & \multicolumn{1}{c|}{0.784}  & \multicolumn{1}{c|}{\textbf{0.819}} & \multicolumn{1}{c|}{0.391}    & \multicolumn{1}{c|}{0.373}          & \multicolumn{1}{c|}{\textbf{0.424}} & \multicolumn{1}{c|}{0.627}    & \multicolumn{1}{c|}{0.593}       & \multicolumn{1}{c|}{\textbf{0.686}}       & \multicolumn{1}{c|}{\textbf{0.820}}    & \multicolumn{1}{c|}{0.794}       &  \multicolumn{1}{c|}{0.808}                                 \\ \hline
\end{tabular}
}
\label{dataset_augmentation}
\begin{tablenotes}
    \item[a] synth. aug. -- augmentation with positive synthetic samples.
\end{tablenotes}
\end{threeparttable}
\end{table*}

\begin{table*}[htbp]
\caption{Convergence Analysis}
\begin{threeparttable}
\centering
\resizebox{\textwidth}{!}{
\begin{tabular}{|c|cccccccccccccccccccccc|}
\hline
\multicolumn{1}{|l|}{Metrics} & \multicolumn{11}{c|}{\textbf{Train Synthetic Dataset - Test Real Dataset (TSTR)}}                                                                                                                                                                                                                                                                                & \multicolumn{11}{c|}{\textbf{Train Real Dataset - Test Synthetic Dataset (TRTS)}}                                                                                                                                                                                                                                                                        \\ \hline
\multirow{3}{*}{\textbf{}}    & \multicolumn{22}{c|}{\textbf{Abnormality}}                                                                                                                                                                                                                                                                                                                                                                                                                                                                                                                                                                                                                                                                                    \\ \cline{2-23} 
                              & \multicolumn{4}{c|}{\textbf{AFIB}}                                                                                         & \multicolumn{4}{c|}{\textbf{CRBBB}}                                                                                                 & \multicolumn{3}{c|}{\textbf{PVC}}                                                             & \multicolumn{4}{c|}{\textbf{AFIB}}                                                                                         & \multicolumn{4}{c|}{\textbf{CRBBB}}                                                                                        & \multicolumn{3}{c|}{\textbf{PVC}}                                                              \\ \cline{2-23} 
                              & \multicolumn{1}{l|}{4k \tnote{a}}    & \multicolumn{1}{l|}{24k}   & \multicolumn{1}{l|}{100k}  & \multicolumn{1}{l|}{1.2m \tnote{b}}           & \multicolumn{1}{l|}{4k}    & \multicolumn{1}{l|}{24k}   & \multicolumn{1}{l|}{80k}            & \multicolumn{1}{l|}{880k}           & \multicolumn{1}{l|}{48k}   & \multicolumn{1}{l|}{576k}  & \multicolumn{1}{l|}{1.632m}         & \multicolumn{1}{l|}{4k}    & \multicolumn{1}{l|}{24k}   & \multicolumn{1}{l|}{100k}  & \multicolumn{1}{l|}{1.2m}           & \multicolumn{1}{l|}{4k}    & \multicolumn{1}{l|}{24k}   & \multicolumn{1}{l|}{80k}            & \multicolumn{1}{l|}{880k}  & \multicolumn{1}{l|}{48k}   & \multicolumn{1}{l|}{576k}           & \multicolumn{1}{l|}{1.632m} \\ \hline
G-mean                        & \multicolumn{1}{c|}{0.507} & \multicolumn{1}{c|}{0.828} & \multicolumn{1}{c|}{0.745} & \multicolumn{1}{c|}{\textbf{0.922}} & \multicolumn{1}{c|}{0.885} & \multicolumn{1}{c|}{0.908} & \multicolumn{1}{c|}{\textbf{0.940}} & \multicolumn{1}{c|}{0.853}          & \multicolumn{1}{c|}{0.561} & \multicolumn{1}{c|}{0.847} & \multicolumn{1}{c|}{\textbf{0.850}} & \multicolumn{1}{c|}{0.541} & \multicolumn{1}{c|}{0.632} & \multicolumn{1}{c|}{0.809} & \multicolumn{1}{c|}{\textbf{0.985}} & \multicolumn{1}{c|}{0.238} & \multicolumn{1}{c|}{0.980} & \multicolumn{1}{c|}{\textbf{1.000}} & \multicolumn{1}{c|}{0.992} & \multicolumn{1}{c|}{0.329} & \multicolumn{1}{c|}{\textbf{0.864}} & 0.808                       \\ \hline
f1-score                      & \multicolumn{1}{c|}{0.308} & \multicolumn{1}{c|}{0.434} & \multicolumn{1}{c|}{0.564} & \multicolumn{1}{c|}{\textbf{0.673}} & \multicolumn{1}{c|}{0.218} & \multicolumn{1}{c|}{0.378} & \multicolumn{1}{c|}{0.640}          & \multicolumn{1}{c|}{\textbf{0.678}} & \multicolumn{1}{c|}{0.121} & \multicolumn{1}{c|}{0.400} & \multicolumn{1}{c|}{\textbf{0.561}} & \multicolumn{1}{c|}{0.163} & \multicolumn{1}{c|}{0.191} & \multicolumn{1}{c|}{0.292} & \multicolumn{1}{c|}{\textbf{0.858}} & \multicolumn{1}{c|}{0.095} & \multicolumn{1}{c|}{0.953} & \multicolumn{1}{c|}{\textbf{0.981}} & \multicolumn{1}{c|}{0.757} & \multicolumn{1}{c|}{0.149} & \multicolumn{1}{c|}{\textbf{0.649}} & 0.621                       \\ \hline
\end{tabular}
}
\label{convergence_analysis}
\begin{tablenotes}
    \item[a] 4k -- model checkpoint for 4000 samples.
    \item[b] 1.2m -- model checkpoint for 1.2 million samples.
\end{tablenotes}
\end{threeparttable}
\end{table*}

\subsubsection{Generative model convergence assessing} \label{exp:convergence}

In Section \ref{methods:model_eval}, we discussed the challenges of evaluating generative models synthesizing ECG signals. These challenges naturally arise during the validation phase in the training process of a diffusion model. Moreover, each of the described metrics requires the generation of synthetic samples, and for proper validation, a statistically significant number of such samples is necessary. Since generating samples with a diffusion model requires a considerable amount of time, several times exceeding the training time, validation becomes computationally infeasible. One method to address this is to accelerate the denoising procedure through distillation \cite{b39} \cite{b40} or directly finding a more efficient way to solve the equation defining the backward process of the model \cite{b41}. We also see this task as a potential direction for future work.

As we do not validate the diffusion model, one of the experiments involves assessing the convergence speed. To achieve this, during training, we save the model weights after every 4000 samples have been processed. We fix the number of samples in opposite to the number of iterations, to obtain checkpoints which independent of the batch size. Subsequently, we select several checkpoints at different training stages and repeat the procedure of obtaining the results of Table \ref{conditioning_comparsion}: generating a synthetic copy of the dataset and computing downstream metrics. The convergence results of the diffusion model are presented in Table \ref{convergence_analysis}.

The experiment results confirm the reasonableness of the described validation challenge. Models generating AFIB and PVC signals showed the best performance at the last selected checkpoint, indicating they have not yet reached a plateau and are under-trained after 1 million viewed samples. In contrast, the diffusion model for CRBBB exhibited better results at 80,000 samples than at 880,000, suggesting its optimal weights may lie within the range of up to 1 million processed samples. These results may also explain the decrease in the TSTR metric for PVC compared to AFIB and CRBBB, as this model simply require more iterations for training. All of this underscores the relevance of the validation challenge for the diffusion model.

\subsection{Qualitative assessment} \label{exp:qualitative_assessment}

Qualitative evaluation involves visually examining the generated ECGs. Interpreting electrocardiograms requires expert knowledge. For this reason, several physicians were involved in the visual assessment process. We do not conduct direct qualitative analysis by ourself and focus the experts' resources on solving two related tasks to obtain quantitative metrics of visual assessment.

The first task involved writing a text report for the proposed dataset, consisted of 200 ECG signals. It included 100 real and 100 synthetic ECGs with the following distribution: 20 samples each of AFIB, CRBBB, IMI, PVC, and Other. The quantitative metric was the accuracy of predicting the generated disease. To do this, text reports were manually reviewed, and for each signal, the corresponding keywords indicating the presence or absence of the abnormality under consideration were highlighted. The obtained results are presented in Table \ref{physicians_classification}. For the synthetic subset, "Other" denotes samples with a label of $0$ in the AFIB generation task, meaning these could be any samples except AFIB. They were chosen to test the proposed conditioning mechanism in terms of generating neutral samples. For the real subset, "Other" was taken as Normal ECG (NORM) with Sinus Rhythm (SR). It can be noticed that the obtained results correlate significantly with the results of Table \ref{conditioning_comparsion}. Indeed, by viewing a physician as an expert trained on real examples for interpreting ECGs, we model the TRTS and TRTR scenarios, in which we obtain comparable metrics with quantitative evaluation.

\begin{table}[htbp]
\caption{Accuracy of physicians classification}
\centering
\begin{tabular}{|c|c|c|c|c|c|}
\hline
          & AFIB & CRBBB & PVC  & IMI  & Other \\ \hline
Real      & 0.95 & 0.95  & 1.00 & 0.80 & 1.00  \\ \hline
Synthetic & 0.95 & 1.00  & 0.85 & 0.70 & 1.00  \\ \hline
\end{tabular}
\label{physicians_classification}
\end{table}

In addition to direct abnormality classification, the text report provides a full picture of the examined ECG. This allows for comparing reports for real and generated samples and highlighting features that indirectly indicate the synthetic nature of the data. Thus, such a comparison allows understanding how well physicians can distinguish real samples from synthetic in an indirect task. The main difference was the mention of reversed chest electrodes and low voltage almost for each generated recording, which can be clearly interpreted as doubts about the real nature of the ECG. Also, in 10 out of 100 synthetic records, there was a direct mention that the signal is artifactual. Thus, indirect comparison clearly indicates that physicians can see the difference between real and generated ECGs.

The second task was the direct classification of generated ECGs. For this, physicians were asked: "Is this a generated recording?". The sample consisted of 380 ECGs, marked by a third physician not participating in the first task. Out of 380 samples, 100 were real, and 280 were synthetic. Real samples were taken from the first task. Synthetic samples included 100 records from the first task and 280 additional records. Additional records included the same pathologies as in the first task in the same proportions. They differed in the convergence of checkpoints from section \ref{convergence_analysis} from which they were generated. As a result, a binary confusion matrix was obtained, presented in Table \ref{distinguish_conf}. It can be noticed almost ideal distinguish of synthetic and real samples by physicians.

\begin{table}[htbp]
\caption{Distinguish confusion matrix}
\centering
\begin{tabular}{|c|c|c|c|}
\hline
True Real & False Synthetic & False Real & True Synthetic \\ \hline
99        & 1               & 1          & 279            \\ \hline
\end{tabular}
\label{distinguish_conf}
\end{table}

Qualitative evaluation results significantly differ from quantitative ones. Despite the good generalization ability of generated examples in terms of TSTR metrics, physicians easily found the generated samples, confirmed by a direct question and an indirect task setup. This demonstrates the need for a comprehensive evaluation of the generative model with both quantitative and qualitative aspects. Good quantitative metrics indicate that the generated ECGs can serve as substitutes for real ones. However, the main motivation for using generative models for ECG synthesis in other studies \cite{b17},\cite{b24},\cite{b25} centers on privacy of medical data. The clear distinction made by physicians between synthetic and real samples highlights a significant challenge in the field: producing synthetic ECGs that are both varied and realistic enough to be indistinguishable from real data. Addressing this gap could be a crucial direction for future research.

\section{Conclusion} \label{Conclusion}

In our study, we focused on generating digital 12-lead 10-second ECGs using diffusion models. We refined the conditioning mechanism of the SSSD-ECG model and obtained SSSD-ECG-nle, which outperforms the previous model in terms of TSTR and TRTS metrics, better distinguishing neutral examples. We conducted extensive quantitative and qualitative evaluations and highlighted their features in assessing the generative model. Each evaluation revealed its limitations in obtaining a robust diffusion model for generating ECGs. For quantitative evaluation, the validation problem was considered from the perspective of model convergence, confirming the relevance of the challenge. For qualitative evaluation, significant expert resources were engaged to obtain reliable metrics. This evaluation revealed a gap in ensuring the reliable privacy of generated ECGs.

This work improves existing solution and adds additional clarity to the directions of future work. To maintain reproducibility and consistent evaluation in the future, we make the results of physician evaluations and generated records publicly available.


\end{document}